\documentclass[preprint,12pt]{elsarticle}

\usepackage{amssymb}

\journal{}

\begin{document}

\begin{frontmatter}



\title{Evolutionary dynamics in financial markets with heterogeneities in strategies and risk tolerance}


\author {Wen-Juan Xu$^a$}
\author{Chen-Yang Zhong$^b$}
\author {Fei Ren$^c$}
\author{Tian Qiu$^d$}
\author{Rong-Da Chen$^e$}
\author{Yun-Xin He$^e$}
\author{Li-Xin Zhong$^{e}$}\ead{zlxxwj@163.com}

\address[label1]{School of Law, Zhejiang University of Finance and Economics, Hangzhou, 310018, China}
\address[label2]{Department of Statistics, Stanford University, Stanford, CA 94305-4065, USA}
\address[label3]{School of Business and Research Center for Econophysics, East China University of Science and Technology, Shanghai, 200237, China}
\address[label4]{School of Information Engineering, Nanchang Hangkong University, Nanchang, 330063, China.}
\address[label5]{School of Finance and Coordinated Innovation Center of Wealth Management and Quantitative Investment, Zhejiang University of Finance and Economics, Hangzhou, 310018, China}

\begin{abstract}
In nature and human societies, the effects of homogeneous and heterogeneous characteristics on the evolution of collective behaviors are quite different from each other. It is of great importance to understand the underlying mechanisms of the occurrence of such differences. By incorporating pair pattern strategies and reference point strategies into an agent-based model, we have investigated the coupled effects of heterogeneous investment strategies and heterogeneous risk tolerance on price fluctuations. In the market flooded with the investors with homogeneous investment strategies or homogeneous risk tolerance, large price fluctuations are easy to occur. In the market flooded with the investors with heterogeneous investment strategies or heterogeneous risk tolerance, the price fluctuations are suppressed. For a heterogeneous population, the coexistence of investors with pair pattern strategies and reference point strategies causes the price to have a slow fluctuation around a typical equilibrium point and both a large price fluctuation and a no-trading state are avoided, in which the pair pattern strategies push the system far away from the equilibrium while the reference point strategies pull the system back to the equilibrium. A theoretical analysis indicates that the evolutionary dynamics in the present model is governed by the competition between different strategies. The strategy that causes large price fluctuations loses more while the strategy that pulls the system back to the equilibrium gains more. Overfrequent trading does harm to one's pursuit for more wealth.

\end{abstract}


\end{frontmatter}


\section{Introduction}
\label{sec:introduction}

In financial markets, the price movement is not only related to the good or bad news for the related company but also to people's understanding of the information and their preferences for trading\cite{stauffer1,lin1}. For a homogeneous population, the same beliefs and strategies may lead to the same buying-selling actions, which may further lead to the occurrence of herding effect and crowding effect\cite{zhang1,zhang2}. For a heterogeneous population, the different beliefs and strategies may lead to the different buying-selling actions, which may further lead to the slow changes of the prices\cite{medo1,chau1,zhong1,wiesinger1,sasidevan1,martino1,wong1}. Understanding the effects of homogeneous and heterogeneous population on the evolution of stock market is quite important for the risk management and the construction of an efficient market\cite{manrique1,cao1, burkholz1,zhang3,wawrzyniak1,biswas1,xie1}.

In the study of the effects of individual preferences on the evolutionary dynamics, a variety of agent-based models have been borrowed to model people's social and economic behaviors\cite{schweitzer1,hadzibeganovic1,hadzibeganovic2,gao1,hart2,lo1,lo2,hod1}, among which the minority game and the majority game are mainly used to simulate people's trading behaviors and explore the possible strategic effects and psychological effects\cite{zhang4,challet2,challet3,challet4,challet5,challet6}. Depending upon the minority game, the effects of response time on the evolution of stock prices have been investigated\cite{mosetti1}. The delayed response to the historic information leads to the lag effect of price fluctuations. Depending upon the majority game, the effects of imitation on the evolution of stock prices have been investigated\cite{alfi1,martino2}. Buying and selling the stocks according to most people's choices lead to the occurrence of herding effects in the stock market. Depending upon the mixed minority game, the majority game and the dollar-game\cite{challet1}, the effects of sophisticated individuals on the evolution of stock prices have been investigates. Different kinds of trader-trader interactions lead to typical stylized facts in the stock markets\cite{gabaix1,jiang1,bianconi1,galla2,barato1,marsili1}.

In the study of the effects of investment strategies on the evolution of stock prices, the pair pattern strategy and the reference point strategy are two typical strategies proposed by zhang et al\cite{zhang5,zhang6,ren1}. Depending upon the pair pattern strategies, people buy and sell the stocks frequently, which leads to the occurrence of a power-law return distribution similar to that in real stock markets\cite{mantegna1,mantegna2,cizeau1,gopikrishnan1,plerou1}. Different from the pair pattern strategies which depend on the history of price movement, the reference point strategy is a myopic strategy, which depends on people's subjective cognition and provides us an anchor to simplify our complex decision-making processes\cite{baker1,shi1}. In real markets, the reference point strategies help us finish the transactions in a simplified way.

In this paper, we incorporate the investors with pair pattern strategies and reference point strategies into a trading model. The role of heterogeneities in investment strategies and risk tolerance in the evolution of stock prices is investigated. The following are our main findings.

(1)The heterogeneities in investment strategies and risk tolerance effectively suppress the price fluctuations. As nearly all the investors have similar investment strategies or risk tolerance, the price has a large fluctuation. For the market with heterogeneous investment strategies and heterogeneous risk tolerance, the price fluctuation becomes moderate.

(2)The competition between different investment strategies is related to the coexistence of different strategies. As compared with the strategy which leads to a large price fluctuation, the strategy which leads to a stable price is more competitive.

(3)The coexistence of the investors with pair pattern strategies and reference point strategies makes some of the investors earn more and the others lose more, over-frequent trading is disadvantageous for the investors to earn more.

The paper is organized as follows. In section 2, the agent-based model with pair pattern strategies and reference point strategies is introduced. In section 3, the numerical results are presented. In section 4, a theoretical analysis is given. In section 5, the conclusions are drawn.

\section{The model}
\label{sec:model}
In the present model, there are two kinds of investors: the investors with pair pattern strategies and the investors with reference point strategies. In the following, we introduce the characteristics of different strategies, the updating of strategies and the evolution of stock prices respectively.

\subsection{\label{subsec:levelA}Pair pattern strategies and reference point strategies}

In the present model, there are two kinds of strategies, pair pattern strategies and reference point strategies, which are used to make one's buying and selling decisions.

The pair pattern strategy space consists of a series of buying-and-selling strategy pairs. A buying strategy or a selling strategy consists of a m-bit long binary array. For example, if an individual i's buying strategy is $S_i^{buy}=(110)$ and his selling strategy is $S_i^{sell}=(101)$, facing the latest history of $M=3$ price changes, (rise-rise-drop), individual i buys a stock on condition that the number of stocks in his hand is below the maximum value of $K_{max}$. Facing the latest history of $M=3$ price changes, (rise-drop-rise), individual i sells a stock on condition that the number of stocks in his hand is above the minimum value of $K_{min}$. Or else, individual i does nothing.

The reference point strategy space consists of a series of expected prices, called reference points $P^{ref}$ in the present model. If the latest price $P$ is lower than an individual j's reference point, $P<P_j^{ref}$, individual j buys a stock with probability $\frac{P_j^{ref}-P}{P}$ on condition that the number of stocks in his hand is below the maximum value of $K_{max}$. If the latest price is higher than individual j's reference point, $P>P_j^{ref}$, individual j sells a stock with probability $\frac{P-P_j^{ref}}{P}$ on condition that the number of stocks in his hand is above the minimum value of $K_{min}$. Or else, individual j does nothing.

\subsection{\label{subsec:levelB}Evolution of investment strategies}

An individual i's pair pattern strategy evolves as follows. Initially, individual i randomly chooses $n_S$ strategies from the pair pattern strategy space. At each time step, he gives each strategy a virtual score, which is obtained as if the strategy were being used and the virtual score changes continuously. He makes his buying or selling decision according the strategy with the highest score.

An individual j's reference point strategy evolves as follows. Initially, an individual j randomly chooses a gene $g_j$ from the range of $g_j\in[0,g^{max}]$, which is kept with no change in the evolutionary process.He randomly chooses a strategy $P^{exp}$ from the range of $P^{exp}\in [\bar Pe^{\frac{-\alpha g_j}{N}},\bar Pe^{\frac{\alpha g_j}{N}}]$, in which $\bar P$ is the averaged price in the latest $\Delta t$ steps and $\alpha$ is a pre-given constant. At each time step, he makes his buying or selling decision according to this strategy. If his strategy deviates from the range of $P^{exp}\in [\bar Pe^{\frac{-\alpha g_j}{N}},\bar Pe^{\frac{\alpha g_j}{N}}]$, he randomly chooses a new strategy $P^{exp}$ from the range of $P^{exp}\in [\bar Pe^{\frac{-\alpha g_j}{N}},\bar Pe^{\frac{\alpha g_j}{N}}]$. Or else, he keeps his strategy. From the above evolutionary mechanism we know that, in the present model, a large gene means that an individual has strong risk tolerance and is quite possible to change his strategy less frequently. A small gene means that an individual has weak risk tolerance and is quite possible to change his strategy more frequently.

\subsection{\label{subsec:levelC}Evolution of stock prices}

After all the individuals have made their buying, selling or doing nothing decision, i.e. $a_i$=+1, -1 or 0, the price is updated according to the equation

\begin{equation}
P(t)=P(t-1)e^{\frac{\alpha A}{N}},
\end{equation}
in which $A=\Sigma^N_{i=1} a_i$ and $\alpha$ is a pre-given constant.

\section{Simulation results and discussions}
\label{sec:results}

In numerical simulations, we firstly examine whether the proposed model can reproduce the stylized facts in real financial markets. Secondly, the coupled effects of pair pattern strategies and reference point strategies on price fluctuations are investigated extensively.

\subsection{\label{subsec:level}Reproduction of the stylized facts}

In the study of real financial markets, the distribution of price returns and the autocorrelation of price returns are usually investigated, which reflect the characteristics of price movement. For a random price movement, the distribution of price returns is close to a Poisson distribution. For an autocorrelated price movement, the distribution of price returns is close to a power-law distribution\cite{gu1,gu2}. Some studies have shown that a power-law distribution and long-range autocorrelation exist in real financial markets. We firstly examine whether the present model can reproduce the characteristics of price movement in real financial markets or not.

\begin{figure}
\includegraphics[width=12cm]{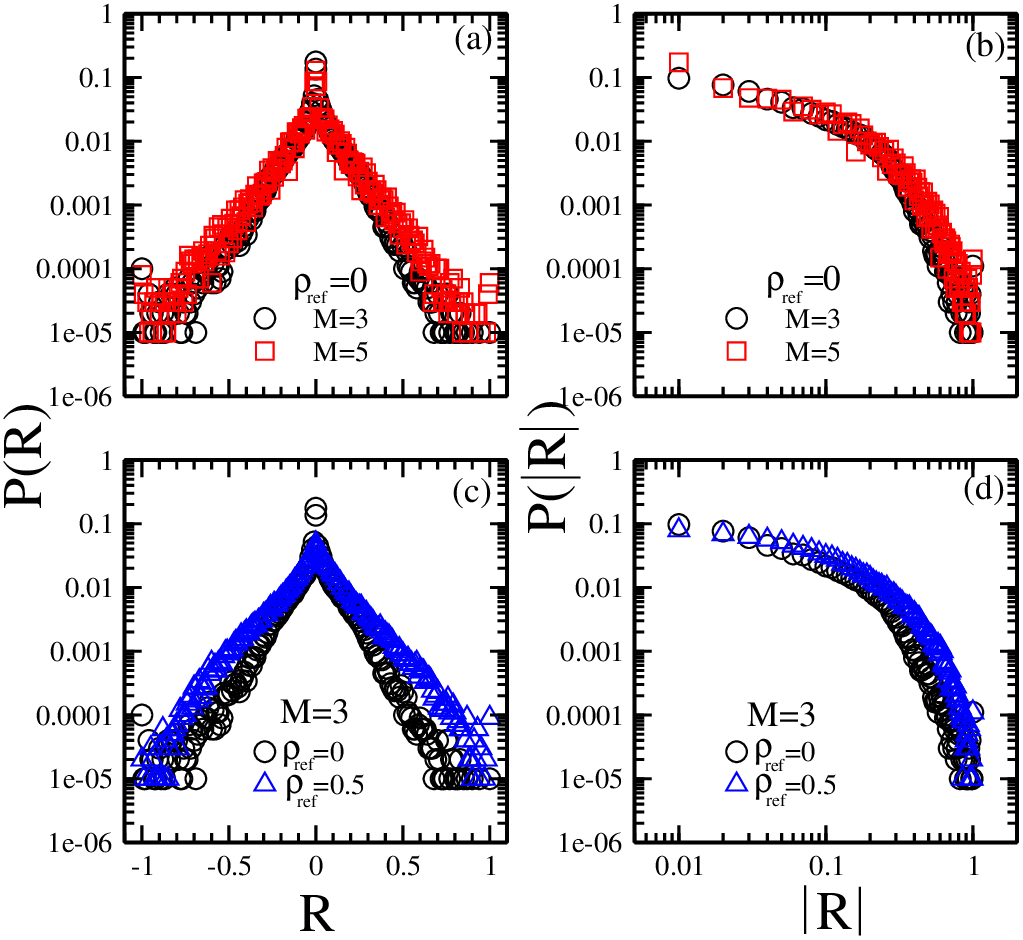}
\caption{\label{fig:epsart}(a)The distribution of price returns $R$ for the ratio of investors with reference point strategies $\rho_{ref}$=0 and the memory size $M$=3 (circles), 5(squares); (b) the distribution of absolute price returns $\vert R \vert$ for $\rho_{ref}$=0 and $M$=3 (circles), 5(squares); (c)The distribution of price returns $R$ for $M$=3 and $\rho_{ref}$=0 (circles), 0.5 (squares); (d) the distribution of absolute price returns $\vert R \vert$ for $M$=3 and $\rho_{ref}$=0 (circles), 0.5 (squares). Other parameters are: the total population $N=5000$, the number of strategies for each investor with pair pattern strategies $n_s=2$, the averaged time $\Delta$ =10, the maximum gene for the investors with reference point strategies $g^{max}$=5000, maximum and minimum stocks for each investor $K_{max}$=1 and $K_{min}$=-1, constant $\alpha$=10.}
\end{figure}

Figure 1 (a) and (b) show that, as all the investors adopt pair pattern strategies, the distribution of price returns is like a power-law distribution, the tail of which is satisfied with the equation $P(\vert R\vert)\sim \vert R \vert^{-\gamma}$. An increase in memory size $M$ leads to a decrease in the exponent $\gamma$. For $M=3$, $\gamma\sim 4.8$. For $M=5$, $\gamma\sim 4$.

Figure 1 (c) and (d) show that, for a given memory size $M$=3, the distribution of price returns is closely related to the ratio of the investors with reference point strategies $\rho_{ref}$. An increase in $\rho_{ref}$ leads to an increase in the exponent $\gamma$. For $\rho_{ref}$=0, $\gamma\sim 4.8$. For $\rho_{ref}$=0.5, $\gamma\sim 8$.

Such results indicate that the distribution of price returns in the present model is closely related to the memory size $M$ and the ratio of the investors with reference point strategies $\rho_{ref}$. Compared with the situation where there is a small $M$ and a small $\rho_{ref}$, a large $M$ and an intermediate $\rho_{ref}$ lead to a broader distribution of price returns.

\begin{figure}
\includegraphics[width=12cm]{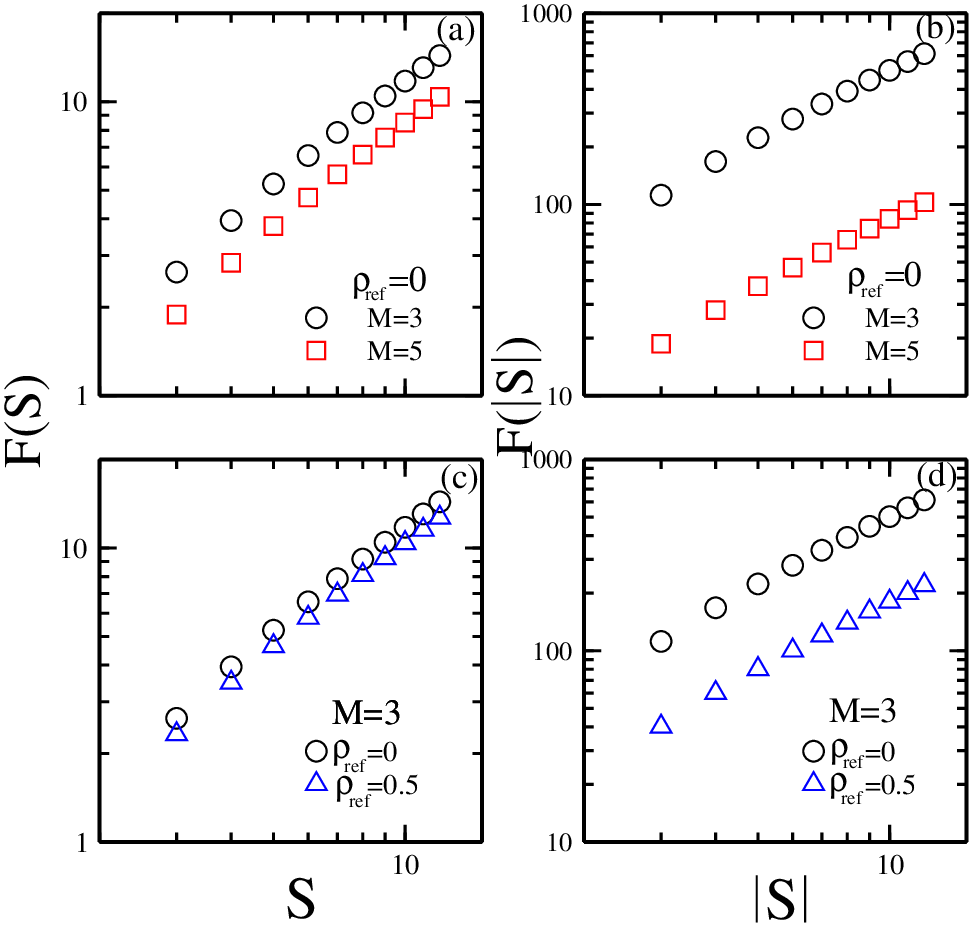}
\caption{\label{fig:epsart}(a)DFA of returns for the ratio of investors with reference point strategies $\rho_{ref}$=0 and the memory size $M$=3 (circles), 5(squares); (b) DFA of absolute returns for $\rho_{ref}$=0 and $M$=3 (circles), 5(squares); (c)DFA of returns for $M$=3 and $\rho_{ref}$=0 (circles), 0.5 (squares); (d) DFA of absolute returns for $M$=3 and $\rho_{ref}$=0 (circles), 0.5 (squares). Other parameters are: total population $N=5000$, number of strategies for each investor with pair pattern strategy $n_s=2$, the averaged time $\Delta$ =10, maximum gene value for the investors with reference point strategy $g^{max}$=5000, maximum and minimum number of stocks for each investor $K_{max}$=1 and $K_{min}$=-1, constant $\alpha$=10.}
\end{figure}

The long-range correlations are usually examined depending upon the detrended fluctuation analysis (DFA). The root mean square of the detrended series $F$ is satisfied with an equation $F(S)\sim S^h$, in which $S$ is the length of local data and $h$ is called hurst exponent. The value of $h$ reflects the moving patterns of price returns. For $h<0.5$, the price returns are satisfied with a short-term correlation. For $h=0.5$, the price returns are satisfied with a random walk. For $h>0.5$, the price returns are satisfied with a long-range correlation.

Figure 2 (a) and (b) show that, for a given ratio of the investors with reference point strategies $\rho_{ref}$=0, the root mean square of the detrended series $F$ is closely related to the memory size $M$. An increase in $M$ leads to an overall decrease in $F$ within the whole range of $S\ge 2$.

Figure 2 (c) and (d) show that,for a given memory size $M$=3, the root mean square of the detrended series $F$ is closely related to the ratio of the investors with reference point strategies $\rho_{ref}$. An increase in $\rho_{ref}$ leads to an overall decrease in $F$ within the whole range of $S\ge 2$. Comparing the results in figure 1(a)-(d), we find that the values of hurst exponent $h$ are nearly the same for different $M$ and $\rho_{ref}$, which is $h\sim 1$.

Such results indicate that the DFA of returns and the DFA of absolute returns are closely related to the memory size $M$ and the ratio of the investors with reference point strategies $\rho_{ref}$. Compared with the situation where there is a small $M$ and a small $\rho_{ref}$, a large $M$ and an intermediate $\rho_{ref}$ lead to an overall decrease in $F$. There exist long-range autocorrelations of price returns for different $M$ and $\rho_{ref}$.

\subsection{\label{subsec:level}Competition between pair pattern strategies and reference point strategies}

In the present model, we are especially concerned about how the heterogeneities in investment strategies affect the price fluctuations. In the following, we firstly examine how the coupling of the memory size of the investors with pair pattern strategies $M$, the maximum gene of the investors with reference point strategies $g^{max}$ and the ratio of the investors with reference point strategies $\rho_{ref}$ affects the time-dependent behaviors of the stock prices.

\begin{figure}
\includegraphics[width=13cm]{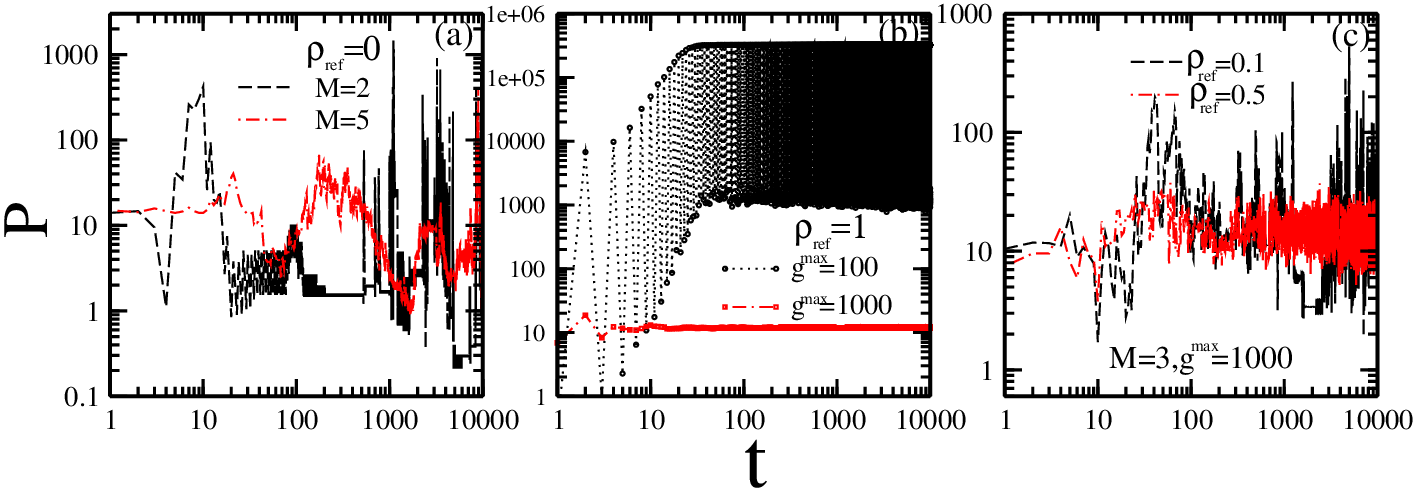}
\caption{\label{fig:epsart}The time-dependent price $P$ (a)for the ratio of the investors with reference point strategies $\rho_{ref}$=0 and the memory size $M=2$ (slashes), 5 (slash dotted lines); (b)for $\rho_{ref}$=1 and the maximum gene of the investors with reference point strategies $g^{max}$=100 (circles),1000 (squares); (c) for $M=3$, $g^{max}=1000$ and $\rho_{ref}$=0.1(slashes), 0.5(slash dotted lines). Other parameters are: total population $N=1000$, the number of strategies for each investor with pair pattern strategies $n_s$=2, the averaged time $\Delta$ =10, maximum and minimum stocks for each investor $K_{max}$=1 and $K_{min}$=-1, constant $\alpha$=10.}
\end{figure}

Figure 3 (a) shows that, as all the investors adopt pair pattern strategies, the price movement is closely related to the memory size $M$. For $M$=2, the price has a large fluctuation. For $M=5$, the price fluctuation becomes ease. In the present model, a large $M$ implies that the investors with pair pattern strategies have more opportunities to choose his personal investment strategies. Given a large $M$ and $S=2$, it is quite possible that all the investors have different investment strategies. The results in figure 3 (a) imply that the heterogeneities in pair pattern strategies suppress the price fluctuations.

Figure 3 (b) shows that, as all the investors adopt reference point strategies, the price movement is closely related to the maximum gene $g^{max}$. For $g^{max}=100$, the price has a zigzag fluctuation. For $g^{max}=1000$, the price becomes stable. In the present model, a large $g^{max}$ implies that the genes of the investors with reference point strategies scatter about a large range of $0<g<g^{max}$. Given a large $g^{max}\sim N$, it is quite possible that all the investors have different genes. The results in figure 3 (b) imply that the heterogeneities in genes suppress the price fluctuations.

Figure 3 (c) shows that, as the investors with pair pattern strategies coexist with the investors with reference point strategies, the price movement is closely related to the ratio of the investors with pair pattern strategies and reference point strategies. An increase in the ratio of the investors with reference point strategies leads to a decrease in the price fluctuations. As compared with the situations in figure 3 (a) and (b), we find that a large $M$ coupled with a large $g^{max}$ can not only inhibit the occurrence of large price fluctuations but also the occurrence of no-trading states.

Such results indicate that the heterogeneities in investment strategies and individual genes has a great impact on the evolution of prices. A homogeneous population promotes price fluctuations while a heterogeneous population suppresses price fluctuations. In the heterogeneous environment where the investors with pair pattern strategies coexist with the investors with reference point strategies, the pair pattern strategies drive the system away from the equilibrium while the reference point strategies draw the system back to the equilibrium. Heterogeneities in the present model are beneficial for stabilizing the stock prices.

\begin{figure}
\includegraphics[width=10cm]{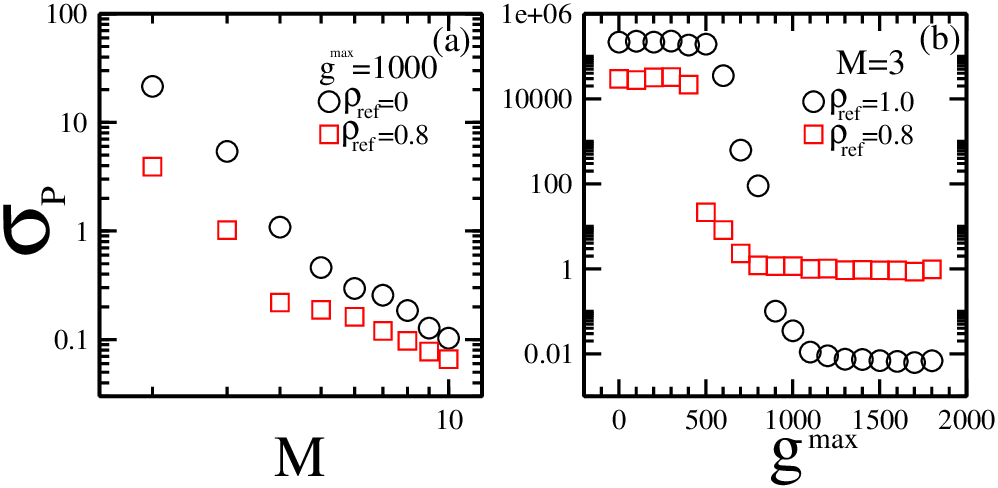}
\caption{\label{fig:epsart}The standard deviation of stock prices $\sigma_P$ (a) as a function of the memory size $M$ for the maximum gene of the investors with reference point strategies $g^{max}=1000$ and the ratio of the investors with reference point strategies $\rho_{ref}$=0 (circles), 0.8 (squares); (b) as a function of $g^{max}$ for $M=3$ and $\rho_{ref}$=1 (circles), 0.8 (squares). Other parameters are: total population $N=1000$, number of strategies for each investor with pair pattern strategy $n_s$=2, the averaged time $\Delta$ =10, maximum and minimum stocks for each investor $K_{max}$=1 and $K_{min}$=-1, constant $\alpha$=10. Final data are obtained by averaging over 100 runs and $10^4$ times after $10^5$ relaxation times in each run.}
\end{figure}

In order to get a clear view of the relationship between the price fluctuations and the heterogeneities in pair pattern strategies and individual genes, in figure 4 (a) and (b) we plot the standard deviation of stock prices $\sigma_P$ as a function of the memory size $M$ and the maximum gene $g^{max}$ respectively.

Figure 4 (a) shows that, as all the investors adopt pair pattern strategies, the price fluctuations are determined by the memory size $M$. As $M$ increases from $M=2$ to $M=10$, the standard deviation of stock prices $\sigma_P$ decreases from $\sigma_P\sim 22$ to $\sigma_P\sim 0.1$. An increase in the ratio of the investors with reference point strategies $\rho_{ref}$ leads to an overall decrease of $\sigma_P$ within the whole range of $2\le M\le 10$. An increase in $\rho_{ref}$ does not affect the changing tendency of $\sigma_P$ vs $M$.

Figure 4 (b) shows that, as all the investors adopt reference point strategies, the price fluctuations are determined by the maximum gene $g^{max}$. There exist two critical points $g^{max}_{c1}\sim 500$ and $g^{max}_{c2}\sim 1100$. As $g^{max}$ increases from $g^{max}=1$ to $g^{max}=500$, $\sigma_P$ keeps a fix value of $2\times 10^5$. As $g^{max}$ increases from $g^{max}=500$ to $g^{max}=1100$, $\sigma_P$ drops quickly from $\sigma_P\sim 2\times 10^5$ to $\sigma_P\sim 0.01$. As $g^{max}$ increases from $g^{max}=1100$ to $g^{max}=1800$, $\sigma_P$ keeps the value of $\sigma_P\sim 0.01$. A decrease in the ratio of the investors with reference point strategies $\rho_{ref}$ leads to a decrease in the critical points $g^{max}_{c1}$ and $g^{max}_{c2}$. Within the range of $1\le g^{max}\le g^{max}_{c2}$, a decrease in $\rho_{ref}$ leads to an overall decrease of $\sigma_P$. Within the range of $g^{max}>g^{max}_{c2}$, a decrease in $\rho_{ref}$ leads to an overall increase of $\sigma_P$. A decrease in $\rho_{ref}$ does not affect the changing tendency of $\sigma_P$ vs $g^{max}$.

Such results indicate that both heterogeneous investment strategies and heterogeneous individual genes have a great impact on the price movement. For the system with either pair pattern strategies or reference point strategies, the heterogeneity suppresses the price fluctuations. For the system with the coexistence of pair pattern strategies and reference point strategies, the price fluctuations may be promoted within some range and may be suppressed within other range.

\begin{figure}
\includegraphics[width=10cm]{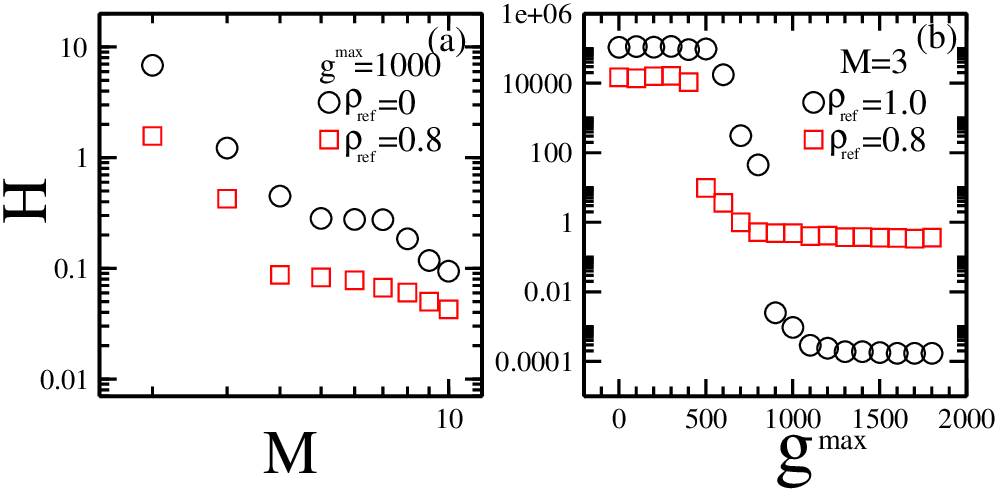}
\caption{\label{fig:epsart}The predictability $H$ of stock prices (a)as a function of the memory size $M$ for the maximum gene of the investors with reference point strategies $g^{max}=1000$ and the ratio of the investors with reference point strategies $\rho_{ref}$=0 (circles), 0.8 (squares); (b)as a function of $g^{max}$ for $M$=3 and $\rho_{ref}$=1 (circles), 0.8 (squares). Other parameters are: total population $N=1000$, number of strategies for each investor with pair pattern strategies $n_s$=2, the averaged time $\Delta$ =10, maximum and minimum stocks for each investor $K_{max}$=1 and $K_{min}$=-1, constant $\alpha$=10. Final data are obtained by averaging over 100 runs and $10^4$ times after $10^5$ relaxation times in each run.}
\end{figure}

In a predictable market, people usually make a deal according to the historic information, which may lead to people's common behaviors. In the present model, even if people have the same historic information, because they have heterogeneous strategies, it is quite possible that their common behaviors may be inhibited. In order to examine whether the reduction of price fluctuations in the present model results from the reduction of the predictability of price movement, in figure 5 (a) and (b) we plot the predictability of stock prices $H$ as a function of the memory size $M$ and the maximum gene $g^{max}$ respectively.

Figure 5 (a) shows that, as all the investors adopt pair pattern strategies, the predictability of stock prices $H$ is closely related to the memory size $M$. As $M$ increases from $M=2$ to $M=10$, $H$ decreases from $H\sim 6.8$ to $H\sim 0.05$. An increase in the ratio of the investors with reference point strategies $\rho_{ref}$ leads to an overall decrease in $H$ within the whole range of $2\le M\le10$. An increase in $\rho_{ref}$ does not affect the changing tendency of $H$ vs $M$. Comparing the results in figure 4 (a) with the results in figure 5 (a), we find that a more predictable market has a larger price fluctuation.

Figure 5 (b) shows that, as all the investors adopt pair pattern strategies, the predictability of stock prices $H$ is closely related to the maximum gene $g^{max}$. There exist two critical points $g^{max}_{c1}\sim 500$ and $g^{max}_{c2}\sim 1100$. As $g^{max}$ increases from $g^{max}=1$ to $g^{max}=500$, $H$ keeps a fix value of $1\times 10^5$. As $g^{max}$ increases from $g^{max}=500$ to $g^{max}=1100$, $H$ drops quickly from $H\sim 1\times 10^5$ to $H\sim 0.002$. As $g^{max}$ increases from $g^{max}=1100$ to $g^{max}=1800$, $H$ keeps the value of $H\sim 0.002$. A decrease in the ratio of the investors with reference point strategies $\rho_{ref}$ leads to a decrease in the critical points $g^{max}_{c1}$ and $g^{max}_{c2}$. Within the range of $1\le g^{max}\le g^{max}_{c2}$, a decrease in $\rho_{ref}$ leads to an overall decrease of $H$. Within the range of $g^{max}>g^{max}_{c2}$, a decrease in $\rho_{ref}$ leads to an overall increase of $H$. A decrease in $\rho_{ref}$ does not affect the changing tendency of $H$ vs $g^{max}$. Comparing the results in figure 4 (b) with the results in figure 5 (b), we find that a more predictable market has a larger price fluctuation.

Such results indicate that the predictability of price movement is closely related to the heterogeneities in investment strategies and individual genes. For the system with either pair pattern strategies or reference point strategies, the heterogeneities are quite possible to make the market become unpredictable, which is similar to the situation in an efficient market. For the system with the coexistence of pair pattern strategies and reference point strategies, the price movement may become more predictable within some range and may become more unpredictable within other range.

\begin{figure}
\includegraphics[width=10cm]{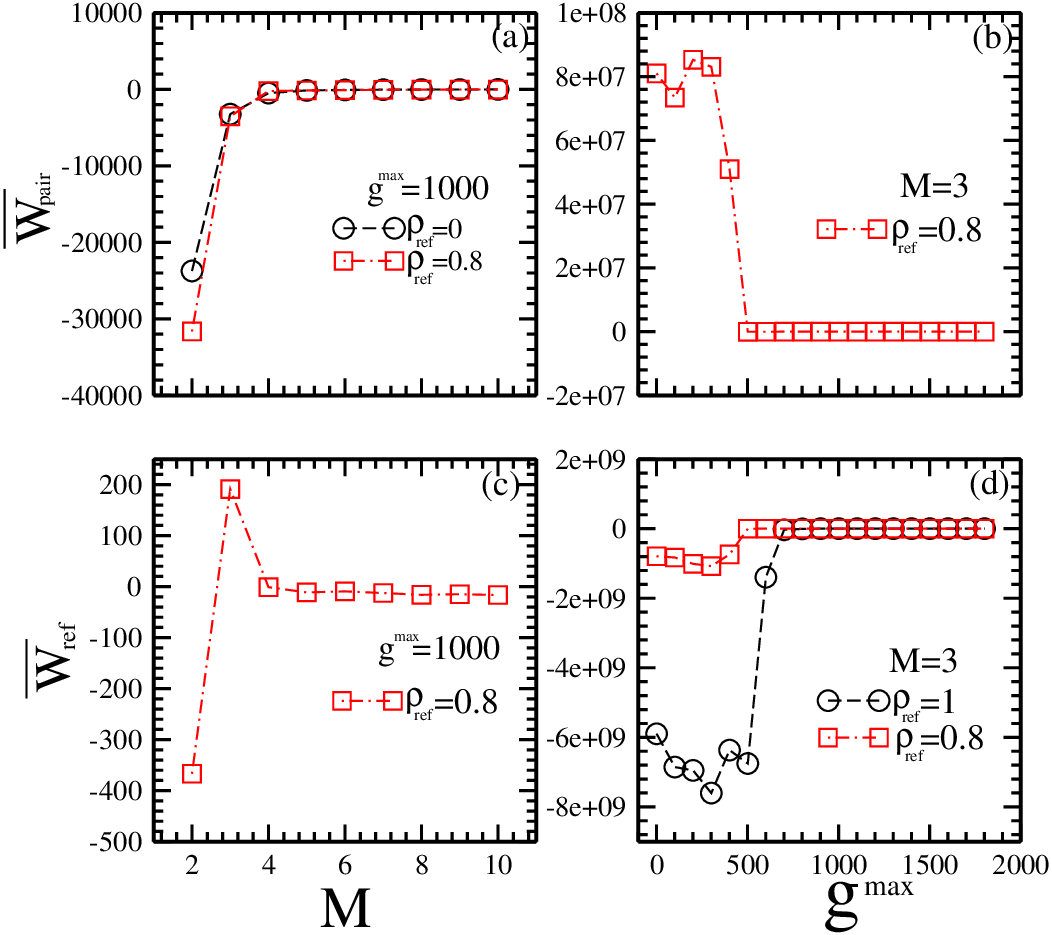}
\caption{\label{fig:epsart}(a)The averaged wealth of the investors with pair pattern strategies $\bar W_{pair}$ as a function of the memory size $M$ for the maximum gene of the investors with reference point strategies $g^{max}=1000$ and the ratio of the investors with reference point strategies $\rho_{ref}$=0 (circles), 0.8 (squares); (b) $\bar W_{pair}$ as a function of $g^{max}$ for $M=3$ and $\rho_{ref}$= 0.8 (squares); (c)the averaged wealth of the investors with reference point strategies $\bar W_{ref}$ as a function of $M$ for $g^{max}=1000$ and $\rho_{ref}$= 0.8 (squares); (d) $\bar W_{ref}$ as a function of $g^{max}$ for $M=3$ and $\rho_{ref}$=1 (circles), 0.8 (squares). Other parameters are: total population $N=1000$,the number of strategies for each investor with pair pattern strategies $n_s$=2, the averaged time $\Delta$ =10, maximum and minimum stocks for each investor $K_{max}$=1 and $K_{min}$=-1, constant $\alpha$=10. Final data are obtained by averaging over 100 runs and $10^4$ times after $10^5$ relaxation times in each run.}
\end{figure}

In order to find a competitive strategy in the present model, in figure 6 (a) - (d) we plot the averaged wealth $\bar W$ of the investors with pair pattern strategies and reference point strategies respectively.

Figure 6 (a) and (c) show that, as all the investors adopt pair pattern strategies, the wealth of the investors with pair pattern strategies $\bar W_{pair}$ is closely related to the memory size $M$. There exists a critical point $M_c=4$. As $M$ increases from $M=2$ to $M=4$, $\bar W_{pair}$ increases quickly from $\bar W_{pair}\sim -2.4\times 10^4$ to $\bar W_{pair}\sim -479$. As $M$ increases from $M=4$ to $M=10$, $\bar W_{pair}$ increases slowly from $\bar W_{pair}\sim -479$ to $\bar W_{pair}\sim -6.3$. An increase in the ratio of the investors with reference point strategies does not affect the critical point $M_c=4$. As $M$ increases from $M=2$ to $M=4$, $\bar W_{pair}$ increases from $\bar W_{pair}\sim -3.2\times 10^4$ to $\bar W_{pair}\sim -224$ while the wealth of the investors with reference point strategies $\bar W_{ref}$ firstly increases from $\bar W_{ref}\sim -367$ to $\bar W_{ref}\sim 191$ and then drops from $\bar W_{ref}\sim 191$ to $\bar W_{ref}\sim -1$. As $M$ increases from $M=4$ to $M=10$, $\bar W_{pair}$ increases from $\bar W_{pair}\sim -224$ to $\bar W_{pair}\sim -20$ while $\bar W_{ref}$ decreases from $\bar W_{ref}\sim -1$ to $\bar W_{ref}\sim -16$. Within the range of $M\le M_c$, the coexistence of pair pattern strategies and reference point strategies is beneficial for the investors with reference point strategies.

Figure 6 (b) and (d) show that, as all the investors adopt reference point strategies, the wealth of the investors with reference point strategies $\bar W_{ref}$ is closely related to the maximum gene $g^{max}$. There exist two critical points $g^{max}_{c1}\sim 500$ and $g^{max}_{c2}\sim 700$. Within the range of $g^{max}<500$, $\bar W_{ref}$ fluctuates within the range of $\bar W_{ref}\sim -7\times 10^9$. As $g^{max}$ increases from $g^{max}=500$ to $g^{max}=700$, $\bar W_{ref}$ increases from $\bar W_{ref}\sim -7\times 10^9$ to $\bar W_{ref}\sim -3\times 10^7$. As $g^{max}$ increases from $g^{max}=700$ to $g^{max}=1800$, $\bar W_{ref}$ increases from $\bar W_{ref}\sim -3\times 10^7$ to $\bar W_{ref}\sim -3.8$.

A decrease in the ratio of the investors with reference point strategies $\rho_{ref}$ leads to a decrease in the critical points $g^{max}_{c1}$ and $g^{max}_{c2}$. For $\rho_{ref}$=0.8, as $g^{max}$ increases from $g^{max}=1$ to $g^{max}=300$, the wealth of the investors with pair pattern strategies $\bar W_{pair}$ fluctuates around $\bar W_{pair}\sim 8\times 10^7$ while the wealth of the investors with reference point strategies $\bar W_{ref}$ fluctuate within the range of $\bar W_{ref}\sim -8\times 10^8$. As $g^{max}$ increases from $g^{max}=300$ to $g^{max}=500$, $\bar W_{pair}$ decreases from $\bar W_{pair}\sim 8\times 10^7$ to $\bar W_{pair}\sim -5\times 10^4$ while $\bar W_{ref}$ increases from $\bar W_{ref}\sim -8\times 10^8$ to $\bar W_{ref}\sim -2\times 10^5$. As $g^{max}$ increases from $g^{max}=500$ to $g^{max}=1800$, $\bar W_{pair}$ increases from $\bar W_{pair}\sim -5\times 10^4$ to $\bar W_{pair}\sim -3\times 10^3$ while $\bar W_{ref}$ increases from $\bar W_{pair}\sim -2\times 10^5$ to $\bar W_{pair}\sim -2\times 10^3$. Within the range of $g^{max}\le g^{max}_{c2}$, the coexistence of pair pattern strategies and reference point strategies is beneficial for the investors with pair pattern strategies.

Comparing the results in Figure 6 (a) and (c) with the results in Figure 6 (b) and (d), we find that the coexistence of the investors with pair pattern strategies and reference point strategies is not always good for both sides. In some cases the investors with reference point strategies may defeat the investors with pair pattern strategies and in other cases the investors with  pair pattern strategies may defeat the investors with reference point strategies.

\section{Theoretical analysis}
\label{sec:analysis}
\subsection{\label{subsec:levelA} Relationship between price fluctuations and heterogeneities in pair pattern strategies and risk tolerance}

\begin{figure}
\includegraphics[width=13cm]{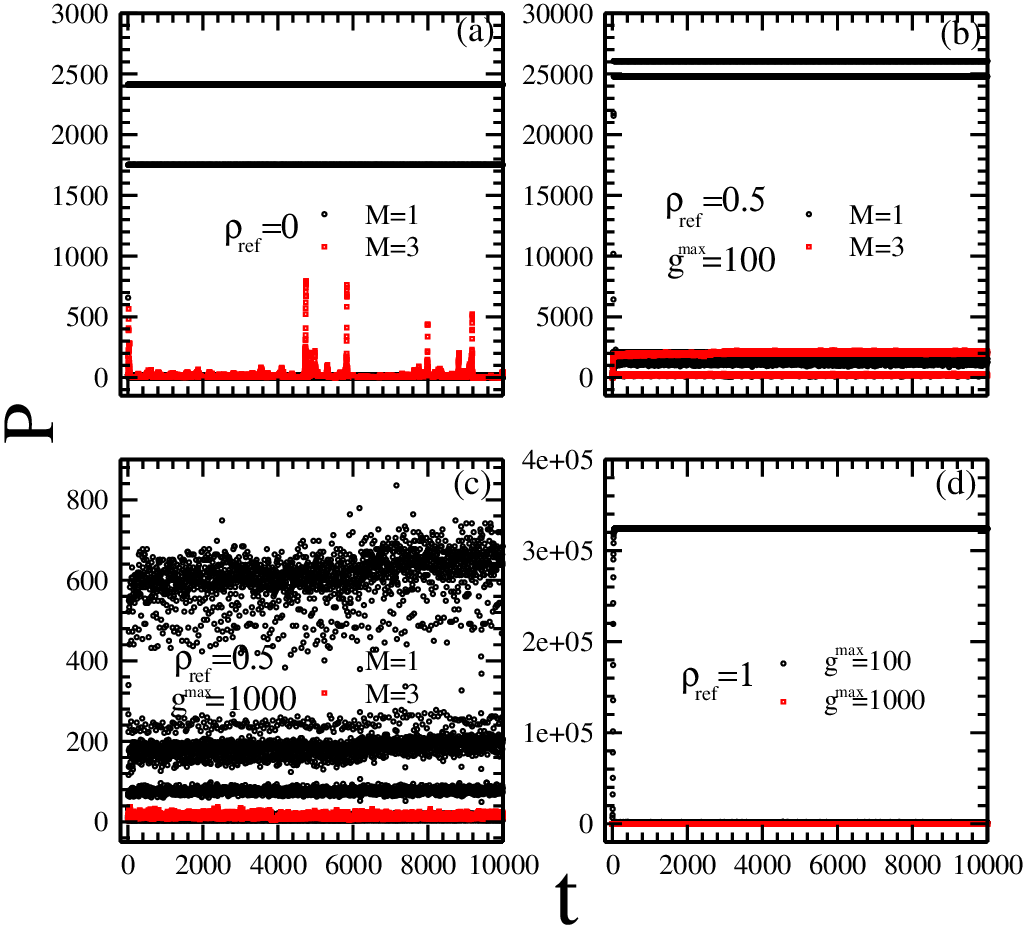}
\caption{\label{fig:epsart}The time-dependent price $P$ for (a) $\rho_{ref}$=0 and $M$=1(circles), 3(squares); (b) $\rho_{ref}$=0.5, $g^{max}=100$ and $M$=1(circles), 3(squares); (c) $\rho_{ref}$=0.5, $g^{max}=1000$ and $M$=1(circles), 3(squares); (d) $\rho_{ref}$=1 and $g^{max}$=100(circles), 1000(squares). Other parameters are: total population $N=1000$, number of strategies for each investors with pair pattern strategy $n_s$=2, the averaged time $\Delta$ =10, maximum and minimum number of stocks for each investor $K_{max}$=1 and $K_{min}$=-1, constant $\alpha$=10.}
\end{figure}

The price fluctuation is determined by the difference in the number of investors buying and selling the stocks, which is satisfied with the equation

\begin{equation}
ln\frac{P(t)}{P(t-1)}={\frac{\alpha \Delta N}{N}}.
\end{equation}

On condition that all the investors adopt pair pattern strategies, for a given population $N$, the heterogeneity in pair pattern strategies is determined by the memory size $M$. For $M=1$, the total number of pair pattern strategies is $n_{pair}=2^M=2$, i.e. $s_1$=(0,1) and $s_2$=(1,0), which means that, facing an increase in the latest price, $s_1$ tells the investors to sell and  $s_2$ tells the investors to buy. Facing a decrease in the latest price, $s_1$ tells the investors to buy and  $s_2$ tells the investors to sell. Therefore, facing a typical change in the latest price, the difference in the number of investors buying and selling the stocks should be

\begin{equation}
\mid\Delta N\mid=\mid N_{s_1}-N_{s_2} \mid,
\end{equation}
in which $N_{s_1}$ and $N_{s_2}$ are the number of investors buying and selling the stocks respectively.

For quite a large $M$, i.e. $M=10$, the number of the combination of $M$ latest price changes is $2^M$=1024. The total number of pair pattern strategies is $n_{pair}=C_{1024}^1C_{1023}^1=1024\times 1023$. For a given $N=1000$, facing a typical combination of the latest $M$ price changes, the difference in the number of investors buying and selling the stocks should be

\begin{equation}
\mid\Delta N\mid\sim 1.
\end{equation}

On condition that all the investors adopt reference point strategies, for a given population $N$, the heterogeneity in reference point strategies is determined by the maximum value of risk tolerance $g^{max}$. Facing the latest price $P(t-1)$, the difference in the number of investors buying and selling the stocks should be

\begin{equation}
\mid\Delta N\mid=\mid N_{P^{ref}>P}-N_{P^{ref}<P} \mid,
\end{equation}
in which $N_{P^{ref}>P}$ and $N_{P^{ref}<P}$ are the number of investors buying and selling the stocks respectively.

For quite a small $g^{max}$, i.e. $g^{max}=1$, all the reference points are within a small range

\begin{equation}
\frac {\bar P}{e}\le P^{ref}\le e\bar P,
\end{equation}
in which $\bar P$ is the averaged value of stock prices in the latest $\Delta t$ steps. For quite a large $g^{max}$, i.e. $g^{max}=N$, all the reference points are within a wide range

\begin{equation}
\frac {\bar P}{e^N}\le P^{ref}\le e^N\bar P.
\end{equation}

Given a typical $\Delta N=N_{P^{ref}>P}-N_{P^{ref}<P}$=2, for $g^{max}=1$, $\Delta N$ is quite possible to become $\Delta N=N_{P^{ref}>P}-N_{P^{ref}<P}$=-2 in the next step because nearly all the investors are within a small range around $P(t-1)$. For $g^{max}=N$, $\Delta N$ is quite possible to become $\Delta N=N_{P^{ref}>P}-N_{P^{ref}<P}$=1 or $\Delta N=N_{P^{ref}>P}-N_{P^{ref}<P}$=0 in the next step because the investors scatter within a wide range around $P(t-1)$.

On condition that the investors with pair pattern strategies coexist with the investors with reference point strategies, the price fluctuation is determined by the coupling of the heterogeneities in investment strategies and risk tolerance. In a heterogeneous population, the existence of the investors with pair pattern strategies helps the investors with reference point strategies away from a no-trading state while the existence of the investors with reference point strategies helps the investors with pair pattern strategies away from a large fluctuation. The price has the characteristics of a random walk which slowly fluctuates around an equilibrium state.

The above analyses indicate that, for a small $M$ and a small $g^{max}$, a zigzag price fluctuation is more possible to occur. For a large $M$ and a large $g^{max}$, a slow price fluctuation like a random walk is more possible to occur. The theoretical analysis is in accordance with the simulation data in figure 7.

\subsection{\label{subsec:levelB} Competition between pair pattern strategies and reference point strategies}

The wealth of the investors with pair pattern strategies and reference point strategies is determined by the difference between the buying price and the selling price.

\begin{equation}
\bar W=\Sigma (P_{sell}-P_{buy}).
\end{equation}
The price change $\Delta P=P(t)-P(t-1)=\Sigma a_{pair}+ \Sigma a_{ref}$. If $\vert\Sigma a_{pair}\vert>>\vert\Sigma a_{ref}\vert$, the price change is determined by the investors with pair pattern strategies. If $\vert\Sigma a_{pair}\vert<<\vert\Sigma a_{ref}\vert$, the price change is determined by the investors with reference point strategies.

\begin{figure}
\includegraphics[width=13cm]{fig8}
\caption{\label{fig:epsart}The averaged wealth of the investors with pair pattern strategies (circles) and reference point strategies (squares) for (a)$M=1$, $g^{max}=100$;(b)$M=1$, $g^{max}=1000$;(c)$M=3$, $g^{max}=100$;(d)$M=3$, $g^{max}=1000$. Other parameters are: total population $N=1000$, the ratio of the investors with reference point strategies $\rho_{ref}$=0.5, number of strategies for each investor with pair pattern strategy $n_s$=2, the averaged time $\Delta$ =10, maximum and minimum number of stocks for each investor $K_{max}$=1 and $K_{min}$=-1, constant $\alpha$=10. Final data are obtained by averaging over 100 runs and $10^4$ times after $10^5$ relaxation times in each run.}
\end{figure}

On condition that only the investors with pair pattern strategies exist, facing a typical combination of $M$ latest price changes, if there are more buyers than sellers, the price increases. The investors buying the stock with a higher price is more than the investors selling the stock with a higher price. Facing another typical combination of $M$ latest price changes, if there are more sellers than buyers, the price decreases. The investors selling the stock with a lower price is more than the investors buying the stock with a lower price. Therefore, buying high and selling low lead to the negative wealth of the investors with pair pattern strategies.

On condition that only the investors with reference point strategies exist, facing the latest price $P(t-1)$, if there are more buyers than sellers, the price increases. The number of investors buying the stock with a higher price is more than the number of investors selling the stock with a higher price. Facing the latest price $P(t-1)$, if there are more sellers than buyers, the price decreases. The investors selling the stocks with a lower price is more than the investors buying the stocks with a lower price. Therefore, buying high and selling low lead to the negative wealth of the investors with reference point strategies.

On condition that the investors with pair pattern strategies coexist with the investors with reference point strategies, the price movement is determined by the strategy governing the moving trend of the price, no matter whether the strategy is pair pattern strategy or reference point strategy. The investors in the minority side gain more than the investors in the majority side.

For a small memory size $M$ and a small maximum gene $g^{max}$, the moving patterns is closely related to the ratio of the investors with pair pattern strategies and reference point strategies. For a small and an intermediate $\rho_{ref}$, because a small $g^{max}$ has a greater impact on the price movement than a small $M$, a large $\rho_{pair}$ has a greater impact on the price movement than a small $\rho_{ref}$. The moving trend of the stock prices is governed by both the investors with pair pattern strategies and reference point strategies. The wealth of the investors with pair pattern strategies is similar to the wealth of the investors with reference point strategies. For a large $\rho_{ref}$, the price movement is governed by the investors with reference point strategies, which means that, if the investors with reference point strategies buy more, the price increases. If the investors with reference point strategies sell more, the price decreases. The majority choice of the investors with reference point strategies determines the moving trend of the price. Therefore, buying high and selling low lead to the negative wealth of the investors with reference point strategies. For the investors with pair pattern strategies, if they have the buying and selling behaviors similar to the investors with reference point strategies, buying high and selling low lead to their negative wealth. If they have the buying and selling behaviors different from the investors with reference point strategies, buying low and selling high lead to their positive wealth. Therefore, compared with the wealth of the investors with reference point strategies, the wealth of the investors with pair pattern strategies is quite possible to be larger than 0.

For a small memory size $M$ and a large maximum gene $g^{max}$, because a large $g^{max}$ has little impact on the price movement, the moving trend of the price is determined by the investors with pair pattern strategies. If the investors with pair pattern strategies buy more, the price increases. If the investors with pair pattern strategies sell more, the price decreases. Therefore, buying high and selling low lead to the negative wealth of the investors with pair pattern strategies. For the investors with reference point strategies, if they have the buying and selling behaviors similar to the investors with pair pattern strategies, buying high and selling low lead to their negative wealth. If they have the buying and selling behaviors different from the investors with pair pattern strategies, buying low and selling high lead to their positive wealth. Therefore, compared with the wealth of the investors with pair pattern strategies, the wealth of the investors with reference point strategies is quite possible to be larger than 0.

For an intermediate memory size $M$ and a small maximum gene $g^{max}$, similar to the situation where there is a small memory size $M$ and a small maximum gene $g^{max}$, the moving patterns is closely related to the ratio of the investors with pair pattern strategies and reference point strategies $\rho_{ref}$. For a small and an intermediate $\rho_{ref}$, because a small $g^{max}$ has a greater impact on the price movement than an intermediate $M$ and a large $\rho_{pair}$ has a greater impact on the price movement than a small $\rho_{ref}$, the moving trend of the price is governed by both the investors with pair pattern strategies and reference point strategies. The wealth of the investors with pair pattern strategies is similar to the wealth of the investors with reference point strategies. For a large $\rho_{ref}$, the price movement is governed by the investors with reference point strategies, which means that, if the investors with reference point strategies buy more, the price increases. If the investors with reference point strategies sell more, the price decreases. The majority choice of the investors with reference point strategies determines the moving trend of the price. Therefore, buying high and selling low lead to the negative wealth of the investors with reference point strategies. For the investors with pair pattern strategies, if they have the buying and selling behaviors similar to the investors with reference point strategies, buying high and selling low lead to their negative wealth. If they have the buying and selling behaviors different from the investors with reference point strategies, buying low and selling high lead to their positive wealth. Therefore, compared with the wealth of the investors with reference point strategies, the wealth of the investors with pair pattern strategies is quite possible to be larger than 0.

For an intermediate memory size $M$ and a large maximum gene $g^{max}$, similar to the situation where there is a small memory size $M$ and a large maximum gene $g^{max}$, because a large $g^{max}$ has little impact on the price movement, the moving trend of the price is determined by the investors with pair pattern strategies. If the investors with pair pattern strategies buy more, the price increases. If the investors with pair pattern strategies sell more, the price decreases. Therefore, buying high and selling low lead to the negative wealth of the investors with pair pattern strategies. For the investors with reference point strategies, if they have the buying and selling behaviors similar to the investors with pair pattern strategies, buying high and selling low lead to their negative wealth. If they have the buying and selling behaviors different from the investors with pair pattern strategies, buying low and selling high lead to their positive wealth. Therefore, compared with the wealth of the investors with pair pattern strategies, the wealth of the investors with reference point strategies is quite possible to be larger than 0.

The above analyses indicate that the strategy that drives the system far away from the equilibrium loses more while the strategy that draws the system back to the equilibrium gains more. The theoretical analysis is in accordance with the simulation data in figure 8.

\section{Summary}
\label{sec:summary}

In an efficient market, the price movement fully reflects good or bad news for the related companies. However, in real societies, in the face of common information, the price movement often depends upon the characteristics of the investors. Homogeneous population and heterogeneous population have an entirely different effect on the price movement.

By incorporating pair pattern strategies and reference point strategies into a trading model, we have investigated the coupled effects of heterogeneous investment strategies and heterogeneous risk tolerance on price movement. In the stock market flooded with the investors with pair pattern strategies, homogeneous investment strategies lead to the occurrence of large price fluctuations. An increase in the heterogeneity in investment strategies leads to a decrease in price fluctuations. In the stock market flooded with the investors with reference point strategies, the dispersion of individual genes determines the characteristics of price fluctuations. As the investors have similar genes, a large price fluctuation is easy to occur. As the individual genes are well diversified, the price fluctuations can be effectively suppressed and the market is quite possible to be stuck in a no-trading state. The coexistence of pair pattern strategies and reference point strategies not only helps the stock market refrain from large fluctuations but also be stuck in a no-trading state. In the stock market flooded with the investors with heterogeneous investment strategies and heterogeneous individual genes, the investors with pair pattern strategies push the stock market away from the equilibrium while the investors with reference point strategies pull the stock market back to the equilibrium.

The role of heterogeneities in price movement is a quite important issue for us to understand the evolutionary dynamics in financial markets. In the future, people's heterogeneous cognitive behaviors will be further considered in the investigation of the evolutionary dynamics of socioeconomic systems. What the differences between a homogeneous environment and a heterogeneous environment are and how these differences affect the evolution of socioeconomic systems are the favorites of ours.

\section*{Acknowledgments}
This work is the research fruits of Social Science Foundation of Zhejiang Province, National Social Science Foundation of China (Grant No. 20BJL147), Humanities and Social Sciences Fund sponsored by Ministry of Education of China (Grant Nos. 19YJAZH120, 17YJAZH067), National Natural Science Foundation of China (Grant Nos. 71371165, 11865009, 71871094, 71631005, 71773105). Thank professor Yi-Cheng Zhang for his suggestions and discussions about the reference point mechanisms.





\bibliographystyle{model1-num-names}



\end{document}